\documentclass[12pt]{article}
\pdfoutput=1

\usepackage{booktabs}
\usepackage{tabulary}
\usepackage{multirow}
\usepackage{rotating}
\usepackage{epsfig}
\usepackage{psfrag}
\usepackage{latexsym}
\usepackage{fancyhdr}
\usepackage{dsfont}
\usepackage{amsmath}
\usepackage{amssymb}
\usepackage{amsfonts}
\usepackage{mathrsfs}
\usepackage{amsthm}
\usepackage{pifont}
\usepackage{dsfont}
\usepackage{multirow}
\usepackage{array}
\usepackage{chngpage}
\usepackage{longtable}
\usepackage{cite}
\usepackage{bbold}
\usepackage{color}
\usepackage{braket}
\usepackage{colordvi}
\usepackage{fancybox}
\usepackage[footnotesize]{caption2}
\usepackage{graphicx}
\usepackage[center,footnotesize,hang]{subfigure}
\usepackage{bbm}
\usepackage{url}
\usepackage[colorlinks, linkcolor=red, anchorcolor=black, citecolor=green]{hyperref}
\usepackage{multirow}
\usepackage{harpoon}
\usepackage{array}
\usepackage{textcomp}  
\usepackage{enumitem}
\usepackage{mathrsfs}  
\newcommand{\PreserveBackslash}[1]{\let\temp=\\#1\let\\=\temp}
\newcolumntype{C}[1]{>{\PreserveBackslash\centering}p{#1}}
\newcolumntype{R}[1]{>{\PreserveBackslash\raggedleft}p{#1}}
\newcolumntype{L}[1]{>{\PreserveBackslash\raggedright}p{#1}}
\allowdisplaybreaks  

\newcommand{\bq}{\begin{eqnarray}}
\newcommand{\nq}{\end{eqnarray}}

\begin{document}
\title{
\begin{flushright}
\hfill\mbox{\small USTC-ICTS-19-17} \\[5mm]
\begin{minipage}{0.2\linewidth}
\normalsize
\end{minipage}
\end{flushright}
{\Large \bf
Neutrino Masses and Mixing from Double Covering of Finite Modular Groups
\\[2mm]}}
\date{}

\author{Xiang-Gan Liu\footnote{E-mail: {\tt
hepliuxg@mail.ustc.edu.cn}},~
Gui-Jun~Ding\footnote{E-mail: {\tt
dinggj@ustc.edu.cn}}
\\*[20pt]
\centerline{
\begin{minipage}{\linewidth}
\begin{center}
{\it \small
Interdisciplinary Center for Theoretical Study and  Department of Modern Physics,\\
University of Science and Technology of China, Hefei, Anhui 230026, China}\\[2mm]
\end{center}
\end{minipage}}
\\[10mm]}
\maketitle
\thispagestyle{empty}

\begin{abstract}

We extend the even weight modular forms of modular invariant approach to general integral weight modular forms. We find that the modular forms of integral weights and level $N$ can be arranged into irreducible representations of the homogeneous finite modular group $\Gamma'_N$ which is the double covering of $\Gamma_N$. The lowest weight 1 modular forms of level 3 are constructed in terms of Dedekind eta-function, and they transform as a doublet of $\Gamma'_3 \cong T'$. The modular forms of weights 2, 3, 4, 5 and 6 are presented. We build a model of lepton masses and mixing based on $T'$ modular symmetry.

\end{abstract}
\newpage

\section{\label{sec:introduction}Introduction}

The standard Model (SM) is well established after the discovery of Higgs boson. The SM has been precisely tested by a great deal of experiments, and it turns out to be a successful theory of electroweak interactions up to TeV scale~\cite{Tanabashi:2018oca}. However, the SM can explain neither the mass hierarchies among quarks and lepton nor the observed drastically different patterns of quark and lepton flavor mixing. The origin of the flavor structure of the quarks and leptons is one of the most important challenges in particle physics. The most promising approach is to appeal to symmetry considerations. The non-abelian discrete flavor symmetry group has been widely explored to explain lepton mixing angles. Discrete flavor symmetry in combination with generalized CP symmetry can give rise to rather predictive models~\cite{Feruglio:2012cw,Holthausen:2012dk,Ding:2013hpa,Feruglio:2013hia,Ding:2013bpa,Chen:2014tpa,Chen:2014wxa,Chen:2015nha,Everett:2015oka}, see~\cite{Yao:2016zev} for a detailed list of references. In particular, the observed flavor mixing patterns of quark and lepton can be explained simultaneously by the same flavor symmetry group in combination with CP symmetry~\cite{Li:2017abz,Lu:2018oxc,Lu:2019gqp}. The flavor symmetry group is usually broken down to different subgroups in the neutrino and charged lepton sectors by the vacuum expectation values (VEVs) of a set of scalar flavon fields. The vacuum alignment results in certain lepton mixing pattern. However, additional dynamics of the flavor symmetry breaking sector together with certain shaping symmetry are generally needed to obtain the desired vacuum alignment. As a consequence, the resulting models look complicated in some sense. Moreover, the leading order predictions of usual discrete flavor symmetry models are generally subject to corrections from higher dimensional operators which involve multiple flavon insertions.

Recently a new approach of modular invariance as flavor symmetry was proposed to solve the flavor problem of SM~\cite{Feruglio:2017spp}. It is notable that the flavon fields could not be needed and the flavor symmetry could be completely broken by the VEV of the modulus $\tau$ in the supersymmetric modular invariant models. The Yukawa couplings transform non-trivially under the finite modular group $\Gamma_N$ and they can be written in terms of modular forms which are function of $\tau$ with specific modular properties. The superpotential of the theory is strongly constrained by the modular invariance and the all higher dimensional operators in the superpotential are completely  determined in the limit of unbroken supersymmetry. Thus the above mentioned drawback of the usual discrete flavor symmetry can be overcome in modular invariant models~\cite{Feruglio:2017spp}.

The finite modular groups $\Gamma_N\equiv\overline{\Gamma}/\overline{\Gamma}(N)$ for $N\leq5$ are isomorphic to permutation groups. The modular forms for modular groups $\Gamma(2)$~\cite{Kobayashi:2018vbk,Kobayashi:2018wkl,Kobayashi:2019rzp}, $\Gamma(3)$~\cite{Feruglio:2017spp,Criado:2018thu,Kobayashi:2018scp,Okada:2018yrn,Novichkov:2018yse}, $\Gamma(4)$~\cite{Penedo:2018nmg,Novichkov:2018ovf} and $\Gamma(5)$~\cite{Novichkov:2018nkm,Ding:2019xna} have been constructed in a variety of ways, and the related phenomenological predictions for neutrino mixing have all been discussed in the literature. The observed quark masses and CKM mixing matrix can be accommodated in modular invariant models~\cite{Kobayashi:2018wkl,Okada:2018yrn,Okada:2019uoy}. A unification of quark and lepton flavors based on the modular symmetry could also be realized in the framework of $SU(5)$ grand unified theory~\cite{deAnda:2018ecu,Kobayashi:2019rzp}.
Besides the applications in flavor problem of SM, the modular-invariance approach have been also applied to radiatively induced neutrino mass models~\cite{Nomura:2019yft,Nomura:2019jxj} and be exploited to construct dark matter model~\cite{Nomura:2019jxj}. Moreover, the modular invariance has been extended to combine with generalized CP symmetry~\cite{Novichkov:2019sqv} such that the models can become more predictive. A formalism of multiple modular symmetries was developed in~\cite{deMedeirosVarzielas:2019cyj}.

So far only even modular forms are considered when constructing modular invariant models. In the present work, we shall extend the modular invariance approach to general integral weight modular forms, i.e., the odd weight modular forms would be included. We find that the basis vectors of the weight $k$ modular space $\mathcal{M}_{k}(\Gamma(N))$ can be decomposed into different irreducible presentations of the homogeneous finite modular groups $\Gamma'_N \equiv \Gamma/\Gamma(N)$, while the frequently studied weight modular forms of even weights transform as irreducible representations of the inhomogeneous finite modular groups $\Gamma_N \equiv \overline{\Gamma}/\overline{\Gamma}(N)$. Notice that $\Gamma'_N$ is the double covering of $\Gamma_N$, yet $\Gamma_N$ is not a subgroup of $\Gamma'_N$. Thus in order to study the odd weight modular forms, one need to consider the finite modular group  $\Gamma'_N$ instead of $\Gamma_N$. The modular forms of level $N$ can be constructed from the tensor products of the lowest weight 1 modular forms. For $N=3$, $\Gamma'_3$ is isomorphic to $T'$ which is the double cover group of $\Gamma_3 \cong A_4$. We construct the modular forms of level 3 up to weight 6 in terms of the Dedekind eta-function. As an example of application of our results, we construct a phenomenologically viable model of neutrino masses and mixing based on $\Gamma'_3\cong T'$.

The rest of this paper is organized as follows. In section~\ref{sec:integral_weight modular_sym}, we briefly review of the formalism of the supersymmetric modular invariant theory, and show that the modular forms of integral weight $k$ and level $N$ transform according to irreducible representations of $\Gamma'_N$. We show the lowest weight 1 modular forms transform in a doublet $\mathbf{2}$ of $T'$ in section~\ref{sec:constructing_Intg_MF}, and modular forms of weight 2, 3, 4, 5, 6 are constructed from the tensor products of the weight 1 modular forms. In section~\ref{sec:model}, we build a modular invariant model with $T'$ symmetry, the weight 3 modular forms enter into the neutrino Yukawa couplings. Section~\ref{sec:conclusion} concludes the paper. We present the $T'$ group theory and the Clebsch-Gordan coefficients in Appendix~\ref{sec:Tp_group}.

\section{\label{sec:integral_weight modular_sym}Modular symmetry and double covering of finite modular group  }

The full modular group $SL(2,\mathbb{Z})$ is the group of 2-by-2 matrices with integral entries and determinant 1~\cite{Bruinier2008The,diamond2005first},\\
\begin{equation}
SL(2,\mathbb{Z})=\left\{\left(\begin{array}{cc}a&b\\c&d\end{array}\right)\bigg|a,b,c,d\in \mathbb{Z},ad-bc=1\right\}\,.
\end{equation}
The modular group $\overline{\Gamma}$ is the linear fraction transformations of the upper half complex plane ${\cal H}=\{\tau\in\mathbb{C}~|~\rm{Im}\,\tau >0\}$, and it has the following form
\begin{equation}
\tau \mapsto \gamma\tau\equiv\frac{a\tau+b}{c\tau+d},\quad \gamma=\begin{pmatrix}
a  &  b  \\
c  &  d
\end{pmatrix}\in SL(2, \mathbb{Z})\,.
\end{equation}
Obviously $\gamma$ and $-\gamma$ lead to the same linear fractional transformation. Therefore the modular group $\overline{\Gamma}$ is isomorphic to the projective special linear group $PSL(2,\mathbb{Z})=SL(2,\mathbb{Z})/\{I, -I\}$, where $I$ is the two-dimensional unit element. It is well-known that the modular group $\bar{\Gamma}$ can be generated by two elements $S$ and $T$~\cite{Bruinier2008The}
\begin{equation}
S: \tau \mapsto -\frac{1}{\tau},\qquad T: \tau \mapsto \tau+1\,,
\end{equation}
which are represented by the following two by two matrices of $SL(2,\mathbb{Z})$
\begin{equation}
S=\left(
\begin{array}{cc}
0 ~&~ 1\\
-1 ~&~ 0
\end{array}
\right),\qquad T=\left(
\begin{array}{cc}
1 ~&~  1\\
0 ~&~ 1
\end{array}
\right)\,.
\end{equation}
It is straightforward to check that the two generators satisfy the following relations
\begin{equation}
\label{eq:muti_rules}S^2=-I,\qquad (ST)^3=I\,.
\end{equation}
Since $I$ and $-I$ are indistinguishable in $PSL(2,\mathbb{Z})$, the generators $S$ and $T$ of $\overline{\Gamma}$ satisfy the famous multiplication rules~\cite{Bruinier2008The}
\begin{equation}
S^2=(ST)^3=\mathds{1}\,,
\end{equation}
where $\mathds{1}$ denotes the identity element of group. The \textit{principal congruence subgroup of level N} for any positive integer $N$ is the subgroup
\begin{equation}
\Gamma(N)=\left\{\left(\begin{array}{cc}a&b\\c&d\end{array}\right)\in SL(2,\mathbb{Z}),~~ \left(\begin{array}{cc}a&b\\c&d\end{array}\right)=\begin{pmatrix}
1  ~&~  0 \\
0  ~&~ 1
\end{pmatrix}~(\text{mod}~N)
\right\}\,,
\end{equation}
which is a infinite normal subgroup of $SL(2,\mathbb{Z})$. Obviously we have $\Gamma(1)\cong SL(2, \mathbb{Z})$ which will be denoted as $\Gamma$ for simplicity of notation in the following. We define $\overline{\Gamma}(N)=\Gamma(N)/\{I, -I\}$ for $N=1, 2$,  while $\overline{\Gamma}(N)=\Gamma(N)$ for $N>2$ because $-I$ doesn't belong to $\Gamma(N)$. The quotient group $\Gamma_N\equiv\overline{\Gamma}/\overline{\Gamma}(N)$ is the inhomogeneous finite modular groups. The group $\Gamma_N$ can be generated by two element $S$ and $T$ satisfying
\begin{equation}
S^2=(ST)^3=T^N=\mathds{1}\,.
\end{equation}
We see that $\Gamma_1$ is a trivial group comprising only the identity element, $\Gamma_2$ is isomorphic to $S_3$. Moreover, the isomorphisms $\Gamma_3\cong A_4$, $\Gamma_4\cong S_4$ and $\Gamma_5\cong A_5$ are fulfilled~\cite{deAdelhartToorop:2011re}. The finite modular group $\Gamma_N$ as flavor symmetry has been widely studied to explain neutrino mixing. In the present work, we shall consider another series of finite group $\Gamma'_N\equiv SL(2,\mathbb{Z})/\Gamma(N)$ which is the double cover of $\Gamma_N$. The group $\Gamma'_N$ can be regarded as the group of two-by-two matrices with entries that are integers modulo $N$ and determinant equal to one modulo $N$, and it is also called $SL(2, Z_N)$ or homogeneous finite modular group in the literature~\cite{deAdelhartToorop:2011re,Schoeneberg1974}. The double cover group $\Gamma'_N$ can be obtained from $\Gamma_N$ by including another generator $\mathbb{R}$ which is related to $-I\in SL(2,\mathbb{Z})$ and commutes with all elements of the $SL(2,\mathbb{Z})$ group, such that the generators $S$, $T$ and $\mathbb{R}$ of $\Gamma'_N$ obey the following relations
\begin{equation}
S^{2}=\mathbb{R},\quad (S T)^{3}=\mathbb{1},\quad T^{N}=\mathbb{1}, \quad \mathbb{R}^{2}=\mathbb{1},\quad \mathbb{R}T=T\mathbb{R}\,.
\end{equation}
It's well known that modular form $f(\tau)$ of weight $k$ and level $N$ is a holomorphic function of the complex variable $\tau$, and  under $\Gamma(N)$ it should transform in the following way
\begin{equation}
\label{eq:def_MF}f\left(\frac{a\tau+b}{c\tau+d}\right)=(c \tau+d)^{k} f(\tau) \quad\text{for}\quad \forall~~\gamma=\begin{pmatrix}
a  &  b \\
c  &  d
\end{pmatrix}\in \Gamma(N)\,,
\end{equation}
where $k\ge 0$ is an integer. The function $f(\tau)$ is required to be holomorphic in ${\cal H}$ and at all the cusps. Obviously we have $-I\in\Gamma(N)$ for $N=1, 2$, using the definition of modular form in Eq.~\eqref{eq:def_MF} for $\gamma=-I$, we can obtain
\begin{equation}
\label{eq:vanishing_cond}f(\tau)=(-1)^{k}f(\tau)\,.
\end{equation}
Therefore $\Gamma(1)$ and $\Gamma(2)$ don't have non-vanishing modular forms with odd weight. However, the group $\Gamma(N)$ for $N>2$ have non-vanishing modular forms with odd weight because $-I \notin \Gamma(N>2)$ and the condition in Eq.~\eqref{eq:vanishing_cond} is not necessary. The modular forms of weight $k$ and level $N$ form a linear space $\mathcal{M}_{k}(\Gamma(N))$, and its dimension is~\cite{Bruinier2008The,schultz2015notes},
\begin{subequations}
\begin{eqnarray}
\label{eq:dime_N2}&&\texttt{dim}\mathcal{M}_{2k}(\Gamma(2))=k+1 ,\quad  N=2,\,k\geq1\,,\\
\label{eq:dimNLa}&&\texttt{dim}\mathcal{M}_{k}(\Gamma(N))=\dfrac{(k-1)N+6}{24}N^2 \prod_{p|N}(1-\dfrac{1}{p^2}), \quad~N>2,\,k\geq 2 \,,
\end{eqnarray}
\end{subequations}
Notice that there is no general dimension formula for weight one modular form, but Eq.~\eqref{eq:dimNLa} is still applicable to the case of $N<6$. The linear space $\mathcal{M}_{k}(\Gamma(N))$ of the modular form has been constructed explicitly~\cite{schultz2015notes}. We list the dimension of $\mathcal{M}_{k}(\Gamma(N))$ and the orders of $\Gamma_N$ and $\Gamma'_{N}$ for $2\leq N \leq5$ in table~\ref{tab:modspace and finite group}.

\begin{table}[t!]
\centering
\begin{tabular}{||c|c||c|c||c|c||}
\hline\hline
$N$&$\texttt{dim}\mathcal{M}_{k}(\Gamma(N)) $ &  $\Gamma_N$ & $|\Gamma_N|$ & $\Gamma'_N$  & $|\Gamma'_N|$  \rule[-2ex]{0pt}{5ex}\\
\hline
2& $k/2 + 1\,(k~~\texttt{even})$ &$S_3$ & 6 & $S_3$  & 6 \rule[-2ex]{0pt}{5ex}\\ \hline
3&$k+1$ &  $A_4$   &  12 & $T'$  & 24  \rule[-2ex]{0pt}{5ex}\\ \hline
4&$2k+1$  & $S_4$ &  24  & $S'_4$ & 48  \rule[-2ex]{0pt}{5ex}\\ \hline
5&$5k+1$  &  $A_5$ & 60 & $A'_5$ & 120  \rule[-2ex]{0pt}{5ex}\\
\hline \hline
\end{tabular}
\caption{The dimension formula $\texttt{dim}\mathcal{M}_{k}(\Gamma(N))$ for $2\leq N\leq 5$, and the order of $\Gamma_N$ and $\Gamma'_{N}$. Notice that $\Gamma'_3$ is isomorphic to the $T'$ group, we denote $\Gamma'_4$ and $\Gamma'_5$ as $S'_4$ and $A'_5$ respectively. The group ID of $T'$, $S'_4$ and $A'_5$ in the computer algebra program \texttt{GAP}~\cite{GAP} are $[24,\, 3]$, $[48,\,30]$ and $[120,\, 5]$ respectively.
\label{tab:modspace and finite group}}
\end{table}

Only modular forms of even weights have been used to build models of quark and lepton flavors so far, we shall extend the formalism of modular invariance to general integral modular forms in the following.

\subsection{Transformation of integral weight modular forms under $\Gamma'_N$}

It has been shown that the modular forms of even weight $2k$ and level $N$ can be decomposed into different irreducible representations of $\Gamma_N$ up to the factor $(c\tau+d)^{2k}$ in~\cite{Feruglio:2017spp}. In this section, we shall show that the modular form $f_i(\tau)$ of $\Gamma(N)$ for $N\ge2$ with integral weight $k$ (odd or even) can be arranged into the irreducible representations of the quotient group $\Gamma'_N\equiv\Gamma/\Gamma(N)$. Let's start by defining the so-called automorphy factor~\cite{diamond2005first}:
\begin{equation}
J_k(\gamma,\,\tau)\equiv(c\tau +d)^k, \quad \gamma=\begin{pmatrix}
a  ~&~ b  \\
c  ~&~ d
\end{pmatrix}\in \Gamma
\end{equation}
Then the definition of weight $k$ modular forms $f_i(\tau)$ of $\Gamma(N)$ in Eq.~\eqref{eq:def_MF} can be rewritten as
\begin{equation}
\label{eq:MF_autF}f_i(h\tau)=J_k(h,\tau)f_i(\tau),\quad h\in\Gamma(N)\,.
\end{equation}
After straightforward calculation, it's easy to show that $J_k(h,\tau)$ satisfies the following properties
\begin{equation}
\label{eq:Jk_property}
\begin{aligned}
&J_k(\gamma_1\gamma_2,\,\tau)=J_k(\gamma_1,\,\gamma_2\tau)J_k(\gamma_2,\,\tau),\quad \gamma_1,\gamma_2 \in \Gamma\,,\\
&J_{k}(\gamma^{-1},\gamma\tau)=J^{-1}_k(\gamma, \tau)\,.
\end{aligned}
\end{equation}
In the following, we shall use the notations $\gamma$ and $h$ to represent a generic element of $\Gamma$ and $\Gamma(N)$ respectively, i.e.,$\gamma \in \Gamma$, $h \in \Gamma(N)$. We denote a multiplet of linearly independent modular forms $f(\tau)\equiv(f_1(\tau),\,f_2(\tau),\,\dots,\,f_n(\tau))^T $ with $n=\texttt{dim}\mathcal{M}_{k}(\Gamma(N))$, and define the function
$F_{\gamma}(\tau)\equiv J^{-1}_k(\gamma,\,\tau) f(\gamma\tau)$. Then we have
\begin{align}
\nonumber
F_{\gamma}(h\tau)&= J^{-1}_k(\gamma,\,h\tau)f(\gamma h\tau) \\
\nonumber
&=J^{-1}_k(\gamma,\,h\tau)f(\gamma h \gamma^{-1} \gamma\tau) \\
\nonumber
&=J^{-1}_k(\gamma,\,h\tau)J_k(\gamma h \gamma^{-1},\,\gamma \tau) f(\gamma \tau) \\
\nonumber
&=J_k(h\gamma^{-1},\,\gamma\tau)f(\gamma\tau)\\
\nonumber&=J_k(h,\,\tau)J^{-1}_k(\gamma,\,\tau)f(\gamma\tau)\\
&=J_k(h,\,\tau)F_{\gamma}(\tau)
\label{F(tau)is modular form}
\end{align}
From Eq.~\eqref{F(tau)is modular form}, we observe that the holomorphic functions $F_{\gamma}(\tau)$ are actually modular forms of $\Gamma(N)$ with weight $k$. Therefore $F_{\gamma}(\tau)$ can be written as linear combinations of $f_i(\tau)$, i.e.
\begin{equation}
\label{eq:rho_comb}F_{\gamma}(\tau)=\rho(\gamma)f(\tau)\,,
\end{equation}
which implies
\begin{equation}
\label{eq:decomp_reducible}f(\gamma\tau)=J_k(\gamma,\,\tau)\rho(\gamma)f(\tau)=(c\tau+d)^k\rho(\gamma)f(\tau)\,.
\end{equation}
Notice that the linear combination matrix $\rho(\gamma)$ in Eq.~\eqref{eq:rho_comb} only depends on the modular transformation $\gamma$. Using Eq.~\eqref{eq:decomp_reducible}, we can obtain
\begin{equation}
\label{eq:fg1g2_fir}f(\gamma_1\gamma_2\tau)=J_k(\gamma_1\gamma_2, \tau)\rho(\gamma_1\gamma_2)f(\tau)\,,
\end{equation}
and
\begin{eqnarray}
\nonumber f(\gamma_1\gamma_2\tau)&=&J_k(\gamma_1,\,\gamma_2\tau)\rho(\gamma_1)f(\gamma_2\tau)\\
\nonumber&=&J_k(\gamma_1,\,\gamma_2\tau)J_k(\gamma_2, \tau)\rho(\gamma_1)\rho(\gamma_2)f(\tau)\\
\label{eq:fg1g2_sec}&=&J_k(\gamma_1\gamma_2, \tau)\rho(\gamma_1)\rho(\gamma_2)f(\tau)\,.
\end{eqnarray}
Comparing Eq.~\eqref{eq:fg1g2_fir} with Eq.~\eqref{eq:fg1g2_sec}, we arrive at the following result,
\begin{equation}
\label{eq:homomorphism}\rho(\gamma_1\gamma_2)=\rho(\gamma_1)\rho(\gamma_2)\,.
\end{equation}
From Eq.~\eqref{eq:decomp_reducible} and the definition of modular from in Eq.~\eqref{eq:MF_autF}, we know
\begin{equation}
f(h\tau)=J_k(h,\tau)f(\tau)=J_k(h,\,\tau)\rho(h)f(\tau),\quad h\in\Gamma(N)\,,
\end{equation}
which leads to
\begin{equation}
\label{eq:id_h}\rho(h)=1,\quad h\in\Gamma(N)\,.
\end{equation}
We conclude that $\rho(h)=1$ for any $h \in \Gamma(N)$. Moreover, because the generators $S$ and $T$ have the following properties
\begin{equation}
S^4\in\Gamma(N),\quad (ST)^3\in\Gamma(N),\quad T^{N}\in\Gamma(N),\quad S^2T=TS^2\,,
\end{equation}
consequently we have
\begin{equation}
\label{eq:rho_GammaNp}\rho^4(S)=\rho^3(ST)=\rho^N(T)=1,\qquad \rho(\mathbb{R})\rho(T)=\rho(T)\rho(\mathbb{R})\,.
\end{equation}
From Eqs.~(\ref{eq:homomorphism}, \ref{eq:id_h}, \ref{eq:rho_GammaNp}), we see that $\rho$ essentially is a linear representation of the quotient group $\Gamma'_N\equiv\Gamma / \Gamma(N)$. Generally speaking the representation $\rho$ is reducible, by Maschke's theorem~\cite{rao2006linear}, each reducible representation of a finite group is completely reducible and it can be decomposed into a direct sum of irreducible unitary representations. As a consequence, by properly choosing basis, $\rho$ can be written into a block diagonal form,
\begin{equation}
\label{dim_relation}\rho \sim \rho_{\mathbf{r_1}} \oplus \rho_{\mathbf{r_2}} \oplus \dots \,,~~~\text{with}~~~ \sum_i\texttt{dim}\,\rho_{\mathbf{r}_i} = \texttt{dim}\mathcal{M}_{k}(\Gamma(N))\,,
\end{equation}
where $\rho_{\mathbf{r}_i}$ denotes an irreducible unitary representation of $\Gamma'_N$. In summary, for a given modular forms space $\mathcal{M}_{k}(\Gamma(N))$, its modular forms can always be organized into some modular multiplets which transform as irreducible unitarity representations $\mathbf{r}_i$ of the double covered modular group $\Gamma'_N$. Namely we can find a basis such that a multiplet of modular forms $f_\mathbf{r}(\tau)\equiv\left(f_1(\tau),\,f_2(\tau),\,\dots\right)^T$ satisfy the following equation
\begin{equation}
\label{eq:MF_irr}f_\mathbf{r}(\gamma\tau)= (c\tau + d)^k\rho_{\mathbf{r}}(\gamma)f_\mathbf{r}(\tau),\quad \gamma \in \Gamma\,.
\end{equation}
In particular, we have
\begin{equation}
\label{eq:decom_ST}f_\mathbf{r}(S\tau)= (-\tau)^k\rho_{\mathbf{r}}(S)f_\mathbf{r}(\tau),\quad f_\mathbf{r}(T\tau)= \rho_{\mathbf{r}}(T)f_\mathbf{r}(\tau)\,.
\end{equation}
In practice, we can find the explicit form $f_{\mathbf{r}}$ by solving Eq.~\eqref{eq:decom_ST}, as shown in section~\ref{sec:constructing_Intg_MF}. Let us now consider two linear fraction transformations $\gamma$ and $S^2\gamma$, where $\gamma$ is representative of an element in $\Gamma'_N$. Although $\gamma$ and $S^2\gamma$ are different elements of $SL(2, \mathbb{Z})$, they induce the same linear fraction transformation $\gamma\tau=S^2\gamma\tau$.
Using Eq.~\eqref{eq:MF_irr} for $\gamma$ and $S^2\gamma$, we can obtain
\begin{equation}\label{eq:gamma_-gamma}
\begin{aligned}
&f_\mathbf{r}(\gamma\tau)= (c\tau + d)^k\rho_{\mathbf{r}}(\gamma)f_\mathbf{r}(\tau)\,, \\
&f_\mathbf{r}(S^2\gamma\tau)= (-1)^k(c\tau + d)^k\rho_{\mathbf{r}}(S^2\gamma)f_\mathbf{r}(\tau)\,,
\end{aligned}
\end{equation}
which yields
\begin{equation}
\rho_{\mathbf{r}}(S^2\gamma)=(-1)^k\rho_{\mathbf{r}}(\gamma)\,.
\end{equation}
Therefore the representation matrix of $\mathbb{R}=S^2$ fulfills
\begin{equation}
\begin{cases}
\rho_\mathbf{r}(\mathbb{R})=\rho_\mathbf{r}(\mathbb{1})=1,\quad \text{for}\quad k ~~\texttt{even}\,, \\
\rho_\mathbf{r}(\mathbb{R})= - \rho_\mathbf{r}(\mathbb{1})=-1,\quad \text{for} \quad k~~\texttt{odd}\,.
\end{cases}
\end{equation}
Therefore $\mathbb{R}$ is represented by a unit matrix in the linear space of even weight modular forms, the modular forms of even weight and level $N$ essentially transform in representations of the projective finite modular group $\Gamma_N$ fulfilling $\rho^2_{\mathbf{r}}(S)=\rho^3_{\mathbf{r}}(ST)=\rho^N_{\mathbf{r}}(T)=1$. For odd weight modular forms, $\mathbb{R}$ is represented by a negative unit matrix, the modular forms of odd weight and level $N$ can be arranged into irreducible representations of $\Gamma'_N$ which is the double covering of $\Gamma_N$.

\subsection{Modular invariant supersymmetic theory}

In this section,we shall briefly review the framework of the modular invariant supersymmetric theory. We shall extend the Yukawa couplings as even weight modular forms in previous work to general integral weight modular forms, and the finite modular group $\Gamma_N$ would be promoted to its doble covering group $\Gamma'_{N}$. Considering the $\mathcal{N}=1$ global supersymmetry, the most general form of the action reads
\begin{equation}
\mathcal{S}=\int d^4 x d^2\theta d^2\bar \theta~ K(\Phi_I,\bar{\Phi}_I; \tau,\bar{\tau})+\int d^4 x d^2\theta~ W(\Phi_I,\tau)+\mathrm{h.c.}\,,
\end{equation}
where $K(\Phi_I,\bar{\Phi}_I; \tau,\bar{\tau})$ is the K$\ddot{\mathrm{a}}$hler potential, and $W(\Phi_I,\tau)$ is the superpotential. $\Phi_I$ is a set of chiral supermultiplets, and it  transforms in a representation $\rho_{I}$ of the quotient group $\Gamma'_{N}$ with a weight $-k_{I}$,
\begin{equation}
\label{eq:modular_transform}\left\{\begin{array}{l}
\tau\to \gamma\tau = \dfrac{a \tau+b}{c \tau+d}\,,\\ \\
\Phi_I\to (c\tau+d)^{-k_I}\rho_I(\gamma)\Phi_I\,,
\end{array} \right.\quad \text{with}\quad \gamma=\begin{pmatrix}
a  &  b  \\
c  &  d
\end{pmatrix}\in\Gamma\,,
\end{equation}
where $\rho_I(\gamma)$ is the unitarity representation matrix of the element $\gamma$ and $k_{I}$ is a generic integer. The requirement that the action $\mathcal{S}$ is invariant under the modular transformation of Eq.~\eqref{eq:modular_transform} entails that the K$\ddot{\mathrm{a}}$hler potential should be invariant
up to a K$\ddot{\mathrm{a}}$hler transformation,
\begin{equation}
K(\Phi_I,\bar{\Phi}_I; \tau,\bar{\tau})\rightarrow K(\Phi_I,\bar{\Phi}_I; \tau,\bar{\tau})+f_K(\Phi_I, \tau)+\bar{f}_K(\bar\Phi_I, \bar\tau)\,.
\end{equation}
An example of K$\ddot{\mathrm{a}}$hler potential invariant under  Eq.~\eqref{eq:modular_transform} up to K$\ddot{\mathrm{a}}$hler transformations is of the following form,
\begin{equation}
K(\Phi_I,\bar{\Phi}_I; \tau,\bar{\tau}) =-h \log(-i\tau+i\bar\tau)+ \sum_I (-i\tau+i\bar\tau)^{-k_I} |\Phi_I|^2~~~,
\end{equation}
where $h$ is a real positive constant. After the modulus $\tau$ gets a vacuum expectation value (VEV), the above K$\ddot{\mathrm{a}}$hler potential gives rise to the following kinetic term for the scalar components $\phi_I$ of the supermultiplets $\Phi_I$ and the modulus field $\tau$,
\begin{equation}
\frac{h}{\langle-i\tau+i\bar\tau\rangle^2}\partial_\mu \bar\tau\partial^\mu \tau+\sum_I \frac{\partial_\mu  \bar{\phi}_I \partial^\mu \phi_I}{\langle-i\tau+i\bar\tau\rangle^{k_I}}\,.
\end{equation}
The kinetic term of $\phi_I$ can be made canonical by rescaling the fields $\phi_I$, and it amounts to a redefinition of the superpotential parameters in a concrete model.

The invariance of the action $\mathcal{S}$ under Eq.~\eqref{eq:modular_transform} requires that the superpotential $W(\Phi_I,\tau)$ should be invariant singlet of the homogeneous finite modular group $\Gamma'_{N}$, and the total weight of $W(\Phi_I,\tau)$ should be vanishing. We can expand $W(\Phi_I,\tau)$ in power series of the supermultiplets $\Phi_I$,
\begin{equation}
W(\Phi_I,\tau) =\sum_n Y_{I_1...I_n}(\tau)~ \Phi_{I_1}... \Phi_{I_n}\,.
\end{equation}
In order to ensure invariance of $W(\Phi_I,\tau)$ under the modular transformation in Eq.~\eqref{eq:modular_transform}, the function $Y_{I_1...I_n}(\tau)$ must transform in the following way,
\begin{equation}
\left\{\begin{array}{l}
\tau\to \gamma\tau =\dfrac{a\tau+b}{c\tau+d}\,,\\ \\
Y_{I_1...I_n}(\tau)\to Y_{I_1...I_n}(\gamma\tau)=(c\tau+d)^{k_Y}\rho_{\mathbf{r}_Y}(\gamma)Y_{I_1...I_n}(\tau)\,,
\end{array}
\right.
\end{equation}
with
\begin{equation}
k_Y=k_{I_1}+...+k_{I_n}\,,\qquad \rho_{\mathbf{r}_Y}\otimes\rho_{I_1}\otimes...\otimes\rho_{I_n} \supset \mathbf{1}\,.
\end{equation}
Here $\rho_{\mathbf{r}_Y}$ is an irreducible representation of $\Gamma'_N$, and $k_Y$ is a generic integer. Previous work on modular invariance focuses on even weight modular forms such that $k_Y$ is assumed to be an even integer. As an example, we shall construct the modular forms of level $N=3$ up to weight 6 in the following. All the integral weight modular forms can be constructed through the tensor products of lowest weight 1 modular
forms.

\section{\label{sec:constructing_Intg_MF}Constructing integral weight modular forms of level $N=3$}
The modular forms of weight $k$ and level $N=3$ expands a linear space $M_k(\Gamma(3))$, and the dimension of $M_k(\Gamma(3))$ is $k+1$.
For the lowest nontrivial weight $k=1$, the dimension is equal to $2$. The whole modular space $M_k(\Gamma(3))$ can be constructed from the Dedekind eta-function. The Dedekind eta-function $\eta(\tau)$ was introduced by Dedekind in 1877 and is defined over the upper half complex plane ${\cal H}=\{\tau\in\mathbb{C}~|~\rm{Im}\,\tau >0\}$ by the equation~\cite{diamond2005first,Bruinier2008The,lang2012introduction},
\begin{equation}
\eta(\tau)=q^{1/24}\prod_{n=1}^\infty \left(1-q^n \right),\qquad q\equiv e^{i 2 \pi\tau}\,.
\end{equation}
The $\eta(\tau)$ function can also written into the following infinite series,
\begin{equation}
\eta(\tau)=q^{1/24}\sum^{+\infty}_{n=-\infty} (-1)^nq^{n(3n-1)/2}\,.
\end{equation}
Under the $S$ and $T$ transformations, $\eta(\tau)$ behaves as~\cite{diamond2005first,Bruinier2008The,lang2012introduction}
\begin{equation}
\eta(\tau+1)=e^{i \pi/12}\eta(\tau),\qquad \eta(-1/\tau)=\sqrt{-i \tau}~\eta(\tau)\,.
\end{equation}
Consequently $\eta^{24}(\tau)$ is a modular form of weight 12.

\subsection{\label{subsec:modular_forms_1}Weights 1 modular forms of level $N=3$ }
The modular space $\mathcal{M}_{k}(\Gamma(3))$ has been explicitly constructed through $\eta$ function as follows~\cite{schultz2015notes}
\begin{equation}
\label{eq:Mk_Gamma3}\mathcal{M}_{k}(\Gamma(3))=\bigoplus_{a+b=k,\,a,b\ge0} \mathbb{C} \frac{\eta^{3a}(3\tau)\eta^{3b}(\tau /3 )}{\eta^k(\tau)}\,.
\end{equation}
We can see that the dimension of $\mathcal{M}_{k}(\Gamma(3))$ is $k+1$. For the lowest nontrivial weight 1 modular forms, we can take the basis vectors to be
\begin{eqnarray}
\nonumber&& \hat{e}_1(\tau)=\frac{\eta^{3}(3\tau)}{\eta(\tau)},\quad \hat{e}_{2}(\tau)=\frac{\eta^{3}(\tau / 3)}{\eta(\tau)}\,.
\end{eqnarray}
The above basis vectors $\hat{e}_1$ and $\hat{e}_2$ are linearly independent, and any modular forms of weight 1 and level $N=3$ can be expressed as a linear combination of $\hat{e}_1$ and $\hat{e}_2$.
Under the action of the generator $T$, $\hat{e}_i$ ($i=1, 2$) transform as
\begin{eqnarray}
\hat{e}_1(\tau)\stackrel{T}{\longmapsto} e^{i2\pi /3}\hat{e}_1(\tau),\qquad \hat{e}_{2}(\tau)\stackrel{T}{\longmapsto} 3(1-e^{i2\pi /3})\hat{e}_1 + \hat{e}_2 \,.
\end{eqnarray}
Similarly we find the following transformation properties under another generator $S$
\begin{eqnarray}
\hat{e}_1(\tau) \stackrel{S}{\longmapsto} 3^{-3/2}(-i\tau)\hat{e}_2(\tau),\qquad \hat{e}_{2}(\tau)\stackrel{S}{\longmapsto} 3^{3/2}(-i\tau)\hat{e}_1(\tau) \,.
\end{eqnarray}
As shown in section~\ref{sec:integral_weight modular_sym}, we can always find a basis in $\mathcal{M}_{k}(\Gamma(N))$ such that a multiplet of modular form $f_\mathbf{r}(\tau)\equiv (f_1(\tau),\,f_2(\tau),\,\dots)^T$ of weight $k$ transform in a irreducible representation $\mathbf{r}$ of $\Gamma'_N$. For the modular forms of weight 1 and level 3, we can start from $\hat{e}_1$ and $\hat{e}_2$ to construct a modular multiplet $Y^{(1)}_{\mathbf{2}}$ transforming as a doublet $\mathbf{2}$ of $\Gamma'_3 \cong T'$:
\begin{equation}
\label{eq:modular_space}
Y^{(1)}_{\mathbf{2}}(\tau)=\begin{pmatrix}
Y_1(\tau) \\
Y_2(\tau)
\end{pmatrix}\,,
\end{equation}
with
\begin{equation}
Y_1(\tau)=\sqrt{2}\,e^{i 7\pi/12}\,\hat{e}_1(\tau),\qquad Y_2(\tau)=\hat{e}_1(\tau)-\frac{1}{3}\hat{e}_2(\tau)\,.
\end{equation}
It is straightforward to check that $Y^{(1)}_{\mathbf{2}}(\tau)$ transforms under $S$ and $T$ as follows
\begin{equation}
\label{eq:MF_decomp_ST}Y^{(1)}_{\mathbf{2}}(-1/\tau)=-\tau\rho_\mathbf{2}(S)Y^{(1)}_{\mathbf{2}}(\tau),\qquad Y^{(1)}_{\mathbf{2}}(\tau+1)=\rho_\mathbf{2}(T)Y^{(1)}_{\mathbf{2}}(\tau)\,,
\end{equation}
where the representation matrices $\rho_{\mathbf{2}}(S)$ and $\rho_{\mathbf{2}}(T)$ are given in table~\ref{tab:irred}. The expression of the $q$-expansion of the doublet modular form $Y^{(1)}_{\mathbf{2}}$ is given by
\begin{eqnarray}
\nonumber Y_1(\tau)&=& \sqrt{2}\, e^{i 7\pi/12} q^{1/3} (1 + q + 2 q^2 + 2 q^4 + q^5 + 2 q^6 +...) ,\\
\label{eq:q_exp_weigh1}Y_2(\tau)&=& 1/3 + 2 q + 2 q^3 + 2 q^4 + 4 q^7 + 2 q^9 + ...\,.
\label{q_expansion}
\end{eqnarray}

\subsection{\label{subsec:modular_forms_higher}Weights 2, 3, 4, 5 and 6 modular forms of level $N=3$ }

There are two ways to construct the higher weight modular forms which can decomposed into different irreducible representations of $T'$. The first way is to follow the same approach as we have constructed the weight 1 modular forms from the original basis of Eq.~\eqref{eq:Mk_Gamma3}, the higher weight modular forms can be found by solving equations of Eq.~\eqref{eq:decom_ST}. The second way is by the tensor products of lower weight modular forms, and the Clebsch-Gordan coefficients of $T'$ in Appendix~\ref{sec:Tp_group} are needed. The two methods are actually equivalent, but in practice, the latter is obviously much easier. Thus we will take the second method in the following.

The weight 2 modular forms can be generated from the tensor products of $Y^{(1)}_{\mathbf{2}}$. The $T'$ contraction rule $\mathbf{2}\otimes\mathbf{2}=\mathbf{3}\oplus\mathbf{1}'$ and they can be arranged into different $T'$ irreducible representations as well:
\begin{equation}
\begin{aligned}
&Y^{(2)}_{\mathbf{1}}=\left(Y^{(1)}_{\mathbf{2}}Y^{(1)}_{\mathbf{2}}\right)_{\mathbf{1'}}=Y_1Y_2 - Y_2Y_1 = 0,\\
&Y^{(2)}_{\mathbf{3}}=\left(Y^{(1)}_{\mathbf{2}}Y^{(1)}_{\mathbf{2}}\right)_{\mathbf{3}}=\left(e^{i\pi/6}Y^2_2,\quad \sqrt{2}e^{i7\pi /12}Y_1Y_2,\quad Y^2_1 \right)^T\,.
\end{aligned}
\end{equation}
There are only three linearly independent weight 2 modular forms which can be arranged into a $T'$ triplet $\mathbf{3}$. This is consistent with fact that the modular space $\mathcal{M}_{2}(\Gamma(3))$ has dimension 3. From Eq.~\eqref{eq:q_exp_weigh1}, we can obtain the $q$-expansion of $Y^{(2)}_{\mathbf{3}}$ as follows,
\begin{eqnarray}
Y^{(2)}_{\mathbf{3}} \equiv \begin{pmatrix}
Y^{(2)}_1(\tau)\\Y^{(2)}_2(\tau)\\Y^{(2)}_3(\tau)\end{pmatrix}=
\begin{pmatrix}
\dfrac{1}{9}e^{i\pi/6}(1 + 12q + 36q^2 + 12q^3 + 84q^4 + 72q^5 +\dots )\\
                                                                       \\[-0.1in]
-\dfrac{2}{3}e^{i\pi/6} q^{1/3}(1 + 7q + 8q^2 + 18q^3 + 14q^4 + 31q^5 +\dots) \\
-2 e^{i\pi/6} q^{2/3}(1 + 2q + 5q^2 + 4q^3 + 8q^4 +\dots)
\end{pmatrix}
\label{q_expansion of Y^(2)}\,.
\end{eqnarray}
It coincides with the $q$-expansion of the weight 2 modular forms of $\Gamma(3)$ in Ref.~\cite{Feruglio:2017spp} up to an overall constant $e^{i\pi/6}/9$. Moreover, we can easily see that the constraint $Y^{(2)2}_2+2Y^{(2)}_1Y^{(2)}_3=0$ is fulfilled in our approach.

Next we can construct the weight 3 modular forms from the tensor products of weight 1 and weight 2 modular forms. Using the Clebsch-Gordan coefficients for the contraction $\mathbf{2}\otimes\mathbf{3}=\mathbf{2}\oplus\mathbf{2}'\oplus\mathbf{2}''$, we obtain
\begin{equation}
\begin{aligned}
&Y^{(3)}_{\mathbf{2}}=\left(Y^{(1)}_{\mathbf{2}}Y^{(2)}_{\mathbf{3}}\right)_{\mathbf{2}}= \left(3e^{i\pi/6}Y_1Y^2_2 ,\quad \sqrt{2}e^{i5\pi/12}Y^3_1-e^{i\pi/6}Y^3_2 \right)^T,\\
&Y^{(3)}_{\mathbf{2'}}=\left(Y^{(1)}_{\mathbf{2}}Y^{(2)}_{\mathbf{3}}\right)_{\mathbf{2'}}=\left(0,\quad 0 \right)^T,\\
&Y^{(3)}_{\mathbf{2''}}=\left(Y^{(1)}_{\mathbf{2}}Y^{(2)}_{\mathbf{3}}\right)_{\mathbf{2''}}=\left(Y^3_1+(1-i)Y^3_2,\quad -3Y_2Y^2_1 \right)^T \,.
\end{aligned}
\end{equation}
Therefore the weight 3 modular forms can be arranged into two doublets $\mathbf{2}$ and $\mathbf{2''}$ of $T'$. In the same fashion, we can construct the weight 4, weight 5, and weight 6 modular forms of $\Gamma(3)$ in turn. Although there are different possible ways to construct tensor products, e.g weight 4 modular forms can be constructed from not only the tensor products of weight 1 and weight 3 modular forms but also the tensor products of two weight 2 modular forms, the final results must be identical up to irrelevant overall factors. The involved algebraic calculations are straightforward although a bit tedious, we just give the nonvanishing and independent modular forms of $\Gamma(3)$ with weights 4, 5 and 6 in following:
\begin{equation}
\begin{aligned}
&\hskip-0.12in Y^{(4)}_{\mathbf{3},I}=\left(Y^{(1)}_{\mathbf{2}}Y^{(3)}_{\mathbf{2}}\right)_{\mathbf{3}} = \left(\sqrt{2}e^{i7\pi/12}Y^3_1Y_2-e^{i\pi/3}Y^4_2,~ -Y^4_1-(1-i)Y_1Y^3_2, ~ 3e^{i\pi/6}Y^2_1Y^2_2 \right)^T,\\
&\hskip-0.12in Y^{(4)}_{\mathbf{1'}}=\left(Y^{(1)}_{\mathbf{2}}Y^{(3)}_{\mathbf{2}}\right)_{\mathbf{1'}}=\sqrt{2}e^{i5\pi/12}Y^4_1-4e^{i\pi/6}Y_1Y^3_2 \,, \\
&\hskip-0.12in Y^{(4)}_{\mathbf{1}}=\left(Y^{(1)}_{\mathbf{2}}Y^{(3)}_{\mathbf{2''}}\right)_{\mathbf{1}}=-4Y^3_1Y_2-(1-i)Y^4_2 \,.
\end{aligned}
\end{equation}
Notice that $Y^{(4)}_{\mathbf{3},II}=\left(Y^{(1)}_{\mathbf{2}}Y^{(3)}_{\mathbf{2''}}\right)_{\mathbf{3}}=- Y^{(4)}_{\mathbf{3},I}$ , namely $Y^{(4)}_{\mathbf{3},II}$ is parallel to $Y^{(4)}_{\mathbf{3},I}$. The weight 5 modular forms can be decomposed into three $T'$ two-dimensional irreducible representations $\mathbf{2}$, $\mathbf{2}'$ and $\mathbf{2}''$ as follows,
\begin{equation}
\begin{aligned}
&\hskip-0.15in Y^{(5)}_{\mathbf{2},I}=\left(Y^{(1)}_{\mathbf{2}}Y^{(4)}_{\mathbf{3},I}\right)_{\mathbf{2}}= \left(2\sqrt{2}e^{i7\pi/12}Y^4_1Y_2+e^{i\pi/3}Y_1Y^4_2,~ 2\sqrt{2}e^{i7\pi/12}Y^3_1Y^2_2+e^{i\pi/3}Y^5_2\right)^T,\\
&\hskip-0.15in Y^{(5)}_{\mathbf{2'},I}=\left(Y^{(1)}_{\mathbf{2}}Y^{(4)}_{\mathbf{3},I}\right)_{\mathbf{2'}}=\left(-Y^5_1+2(1-i)Y^2_1Y^3_2,~ -Y^4_1Y_2+2(1-i)Y_1Y^4_2 \right)^T,\\
&\hskip-0.15in Y^{(5)}_{\mathbf{2''}}=\left(Y^{(1)}_{\mathbf{2}}Y^{(4)}_{\mathbf{3},I}\right)_{\mathbf{2''}}=\left(5e^{i\pi/6}Y^3_1Y^2_2-(1-i)e^{i\pi/6}Y^5_2,~ -\sqrt{2}e^{i5\pi/12}Y^5_1-5e^{i\pi/6}Y^2_1Y^3_2 \right)^T \,.
\end{aligned}
\end{equation}
There are another two possible tensor products of weight 5,
\begin{equation}
\begin{aligned}
&Y^{(5)}_{\mathbf{2},II}=\left(Y^{(1)}_{\mathbf{2}}Y^{(4)}_{\mathbf{1}}\right)_{\mathbf{2}}=[-4Y^3_1Y_2-(1-i)Y^4_2](Y_1, Y_2)^{T}\,,\\
&Y^{(5)}_{\mathbf{2'},II}=\left(Y^{(1)}_{\mathbf{2}}Y^{(4)}_{\mathbf{1'}}\right)_{\mathbf{2'}}=[\sqrt{2}e^{i5\pi/12}Y^4_1-4e^{i\pi/6}Y_1Y^3_2](Y_1, Y_2)^{T}\,.
\end{aligned}
\end{equation}
However they are parallel to $Y^{(5)}_{\mathbf{2},I}$ and $Y^{(5)}_{\mathbf{2'},I}$ respectively because the constraints $Y^{(5)}_{\mathbf{2},II}=(1-i)e^{i2\pi/3}Y^{(5)}_{\mathbf{2},I}$ and $Y^{(5)}_{\mathbf{2'},II}=-(1-i) e^{i2\pi/3}Y^{(5)}_{\mathbf{2'},I}$  are fulfilled. Finally we give the weight 6 modular forms of level 3,
\begin{eqnarray}
\nonumber Y^{(6)}_{\mathbf{3},I}&=& \left(Y^{(1)}_{\mathbf{2}}Y^{(5)}_{\mathbf{2},I}\right)_{\mathbf{3}}\\ \nonumber &=& \left(-2(1-i)Y^3_1Y^3_2+iY^6_2,~ -4e^{i\pi/6}Y^4_1Y^2_2-(1-i)e^{i\pi/6}Y_1Y^5_2,~ 2\sqrt{2}e^{i7\pi/12}Y^5_1Y_2+e^{i\pi/3}Y^2_1Y^4_2\right)^T,\\
\nonumber Y^{(6)}_{\mathbf{3},II}&=&\left(Y^{(1)}_{\mathbf{2}}Y^{(5)}_{\mathbf{2'},I}\right)_{\mathbf{3}}\\
\nonumber &=&\left(-Y^6_1+2(1-i)Y^3_1Y^3_2,~ -e^{i\pi/6}Y^4_1Y^2_2+2(1-i)e^{i\pi/6}Y_1Y^5_2,~ 4e^{i\pi/3}Y^2_1Y^4_2-(1+i)e^{i\pi/3}Y^5_1Y_2\right)^T,\\
Y^{(6)}_{\mathbf{1}}&=&\left(Y^{(1)}_{\mathbf{2}}Y^{(5)}_{\mathbf{2''}}\right)_{\mathbf{1}}=(1-i)e^{i\pi/6}Y^6_2-(1+i)e^{i\pi/6}Y^6_1-10e^{i\pi/6}Y^3_1Y^3_2 \,.
\end{eqnarray}
Notice that $Y^{(6)}_{\mathbf{1'}}=\left(Y^{(1)}_{\mathbf{2}}Y^{(5)}_{\mathbf{2},I} \right)_{\mathbf{1'}}=0$,~$Y^{(6)}_{\mathbf{1''}}=\left(Y^{(1)}_{\mathbf{2}}Y^{(5)}_{\mathbf{2'},I} \right)_{\mathbf{1''}}=0$ and $Y^{(6)}_{\mathbf{3},III}=\left(Y^{(1)}_{\mathbf{2}}Y^{(5)}_{\mathbf{2''}}\right)_{\mathbf{3}}=-Y^{(6)}_{\mathbf{3},I}-Y^{(6)}_{\mathbf{3},II}$, namely $Y^{(6)}_{\mathbf{3},III}$ is not independent from $Y^{(6)}_{\mathbf{3},I}$, and $Y^{(6)}_{\mathbf{3},II}$. We can easily verify that the number of non-vanishing independent modular forms are satisfy the dimension relation Eq.~\eqref{dim_relation} for each integral weight. Moreover, we observe that the modular forms of odd weights transform as  two-dimensional representations $\mathbf{2}$, $\mathbf{2}'$ and $\mathbf{2}''$ of $T'$, and the modular forms of even weights transform according to the $T'$ representations $\mathbf{3}$, $\mathbf{1}$, $\mathbf{1}'$ and $\mathbf{1}''$ which are identical with the representations of $A_4$ in our basis. In comparison with $A_4$ modular flavor symmetry, the odd weight modular forms provide new opportunity for model building. We shall built an example model in which the modular forms are involved in next section.

\section{\label{sec:model}A benchmark model with $\Gamma_3\cong T'$ modular symmetry }

In order to show how the odd weight modular form may play a role in determining lepton masses and flavor mixing, we shall build a modular invariant flavor model with the $\Gamma'_3\cong T'$ symmetry. We assign the three generations of the left-handed lepton doublet $L$ to $T'$ triplet $\mathbf{3}$, while the three right-handed charged lepton $e^{c}$, $\mu^c$ and $\tau^c$ transform as $\mathbf{1}$, $\mathbf{1}''$ and $\mathbf{1}'$ respectively. The neutrino masses are generated from the type-I seesaw mechanism and we only introduce two right-handed neutrinos $N^c$ which are embedded into a $T'$ doublet $\mathbf{2}$. We adopt a supersymmetric context, the two Higgs doublets $H_{u,d}$ are invariant under $T'$. We assume that the modular weights of $L$ and $N^c$ are 2 and 1 respectively, while the right-handed charged leptons $e^{c}$, $\mu^c$, $\tau^c$ and the Higgs doublet $H_{u,d}$ are of zero weight. The symmetry assignments to the
minimal supersymmetric Standard Model (MSSM) fields as well as right-handed neutrinos are summarized in table~\ref{tab:generalmod}.

\begin{table}[t!]
\centering
\begin{tabular}{|c|c|c|c|c|c|c|c|}\hline \hline
 &$N^c$& $e^c$  & $\mu^c$ & $\tau^c$& $L$& $H_{u}$ & $H_{d}$\rule[1ex]{0pt}{1ex}\\ \hline
$SU(2)_L\times U(1)_Y$ &$(1,0)$& $(1,1)$& $(1,1)$& $(1,1)$&$(2, -1/2)$& $(2, 1/2)$  & $(2, -1/2)$ \rule[1ex]{0pt}{1ex}\\
\hline
$\Gamma'_3\cong T'$&$\mathbf{2}$& $\mathbf{1}$ & $\mathbf{1''}$ & $\mathbf{1'}$ & $\mathbf{3}$& $\mathbf{1}$ & $\mathbf{1}$ \rule[1ex]{0pt}{1ex}\\
\hline
$k_I$&$1$& $0$& $0$& $0$& $2$& $0$ & $0$ \rule[1ex]{0pt}{1ex}\\
\hline\hline
\end{tabular}
\caption{\label{tab:generalmod}The transformation properties of the MSSM chiral superfields under Standard Model gauge group $SU(2)_L\times U(1)_Y$ and under $T'$ modular symmetry, $k_I$ refers to the modular weights in the modular transformation.}
\end{table}

In our setting, the modular invariant superpotentials for charged lepton sector and neutrino masses can be written as
\begin{eqnarray}
\label{eq:superpotential_We}\mathcal{W}_e &=& \alpha e^cH_d(L\,Y^{(2)}_\mathbf{3})_\mathbf{1}\, +\, \beta \mu^cH_d(L\,Y^{(2)}_\mathbf{3})_\mathbf{1'}+ \gamma \tau^cH_d(L\,Y^{(2)}_\mathbf{3})_\mathbf{1''}\,, \\
\label{eq:superpotential_Wnu}\mathcal{W}_{\nu}&=& g_1((N^c\,L)_\mathbf{2''}Y^{(3)}_\mathbf{2})_\mathbf{1}H_u +g_2((N^c\,L)_\mathbf{2}Y^{(3)}_\mathbf{2''})_\mathbf{1}H_u +\Lambda(N^c\,N^c\,Y^{(2)}_\mathbf{3})_\mathbf{1}\,,
\end{eqnarray}
where $\alpha$, $\beta$, $\gamma$, $g_1$ and $g_2$ are general complex constants, and $\Lambda$ denotes the cutoff scale of the model. Notice that weight 3 modular forms enter into the model through the neutrino Yukawa coupling terms while only even weight modular forms are used. After electroweak symmetry breaking, the superpotential $\mathcal{W}_e$ leads to the following charged lepton mass matrix
\begin{equation}
\mathcal{M}_E= \left(
\begin{array}{ccc}
e^{i\pi/6}\alpha Y^2_2 ~&~ \alpha\, Y^2_1 ~&~\sqrt{2}e^{i7\pi/12}\alpha Y_1Y_2\\
\sqrt{2}e^{i7\pi/12}\beta Y_1Y_2 ~&~  e^{i\pi/6}\beta Y^2_2  ~&~ \beta Y^2_1\\
\gamma Y^2_1 ~&~ \sqrt{2}e^{i7\pi/12}\gamma Y_1Y_2 ~&~   e^{i\pi/6}\gamma Y^2_2
\end{array}
\right)v_d\,,
\end{equation}
where $v_d = \langle H^0_d\rangle$. The phases of the parameters $\alpha$, $\beta$ and $\gamma$ can be absorbed into the right-handed charged lepton fields such that they can be taken to be real without loss of generality. Tye values of $\alpha$, $\beta$ and $\gamma$ are fixed by the charged lepton masses $m_e$, $m_{\mu}$ and $m_{\tau}$ for given modulus $\tau$. Using the Clebsch-Gordan coefficients of $T'$ in
Appendix~\ref{sec:Tp_group}, we can read off from Eq.~\eqref{eq:superpotential_Wnu} the modular invariant Dirac neutrino mass matrix and the right-handed Majorana neutrino mass matrix,
\begin{eqnarray}
\nonumber \mathcal{M}_D&=&
\left(\begin{array}{cc}
-3g_2Y^2_1Y_2 ~&~ (-2g_1+g_2)Y^3_1+\sqrt{2}e^{-i\pi/4}(g_1+g_2)Y^3_2\\
3\sqrt{2}e^{i7\pi/12}g_1Y_1Y^2_2 ~&~ 3\sqrt{2}e^{i7\pi/12}g_2 Y^2_1Y_2\\
-\sqrt{2}e^{i5\pi/12}(g_1+g_2)Y^3_1+e^{i\pi/6}(g_1-2g_2)Y^3_2 ~&~ -3e^{i\pi/6}g_1Y_1Y^2_2
\end{array}
\right) v_u\,,\\[1.0 cm]
\mathcal{M}_N&=&\dfrac{e^{i7\pi/12}}{\sqrt{2}}
\left(
\begin{array}{cc}
2Y_1Y_2 ~&~ Y^2_1\\
Y^2_1   ~&~ \sqrt{2}e^{-i\pi/4}Y^2_2\\
\end{array}
\right)\Lambda\,,
\end{eqnarray}
with $v_u=\langle H^0_u\rangle$. The effective light neutrino mass matrix given by the type-I seesaw formula,
\begin{equation}
\mathcal{M}_\nu=-\mathcal{M}_D \mathcal{M}^{-1}_N \mathcal{M}^T_D\,.
\label{eq:mnufactor}
\end{equation}
We see that the light neutrino mass matrix $\mathcal{M}_\nu$ only depends on one complex parameters $g_2/g_1$ and the modulus $\tau$ besides the overall factor $g^2_1v^2_u/\Lambda$ which controls the absolute scale of neutrino masses. The vacuum expectation value of the modulus $\tau$ is the only source of flavor symmetry breaking in modular invariance theory. We treat $\tau$ as free parameter in the upper complex plane, and a comprehensive numerical scan over the input parameters is performed. We find that good agreement with experimental data can be achieved for inverted ordering neutrino mass spectrum at the point,
\begin{equation}
\begin{gathered}
\tau=-0.3998+1.1688i\,,\quad \beta/\alpha=3435.14\,, \quad \gamma/\alpha=200.128\,,\\
g_2/g_1=1.625+0.084i\,,\quad  g_1^2v_u^2/\Lambda=0.326 \text{ eV}\,,\quad \alpha v_d = 4.419 \text{MeV}\,,
\end{gathered}
\end{equation}
which gives rise to the following values of observables,
\begin{equation}
\label{eq:mixpara}
\begin{gathered}
m_e/m_{\mu}=0.00479\,,\quad
m_{\mu}/m_{\tau}=0.0561\,, \\
\sin^2\theta_{12}=0.3122\,,\quad
\sin^2\theta_{13}=0.0225\,,\quad
\sin^2\theta_{23}=0.5708,\\
\delta_{CP}/\pi=1.313,\quad
\alpha_{21}/\pi=1.259\,,\\
m_1=0.0491\text{ eV},\quad
m_2=0.0499\text{ eV},\quad
m_3=0\text{ eV}\,.
\end{gathered}
\end{equation}
We see that the lightest neutrino is massless $m_3=0$ since two right-handed neutrinos are introduced in the model. As a consequence, the Majorana phses $\alpha_{31}$ is unphysical. It is remarkable that all the three lepton mixing angles and neutrino mass squared differences are in the experimentally preferred $1\sigma$ range~\cite{Esteban:2018azc}. The Dirac CP phase is predicted to be $\delta_{CP}=1.313\pi$ which falls in the $3\sigma$ allowed region~\cite{Esteban:2018azc}. We can infer from Eq.~\eqref{eq:mixpara} that the sum of neutrino masses is $\sum_i m_i=0.099 \text{ eV}$ which is compatible with the latest Planck result on neutrino mass sum $\sum_{i}m_{i}<0.12 \text{ eV}-0.54 \text{ eV}$ at $95/\%$ confidence level~\cite{Aghanim:2018eyx}. Furthermore, the effective Majorana mass $|m_{ee}|$ of neutrinoless double decay is determined to be $|m_{ee}|=0.0252\text{ eV}$, it is testable in future experiments of neutrinoless double beta decay.

\section{\label{sec:conclusion}Conclusion}

In the modular symmetry approach to neutrino masses and mixing~\cite{Feruglio:2017spp}, only even weight modular forms are considered in the literature at present. In this paper, we have extended the framework of modular invariance to include odd weight modular forms. We show that one can always find a basis such that the multiplets of integral weight modular form of level $N$ transform according to different irreducible representations of the homogeneous finite modular group $\Gamma'_N \equiv \Gamma/\Gamma(N)$, while the frequently studied weight modular forms of even weights can be decomposed into irreducible representations of the inhomogeneous finite modular groups $\Gamma_N \equiv \overline{\Gamma}/\overline{\Gamma}(N)$~\cite{Feruglio:2017spp}. Notice that $\Gamma'_N$ is the double covering of $\Gamma_N$, it has twice as many elements as $\Gamma_N$, but $\Gamma_N$ is not a subgroup of $\Gamma'_N$. Therefore to study the contribution of the odd weight modular forms, one should consider the finite modular group $\Gamma'_N$ instead of $\Gamma_N$ as flavor symmetry. The multiplets of higher weight modular forms can be constructed from the tensor products of the lowest non-trivial weight 1 modular forms.

As a demonstration example, we have studied the modular symmetry $\Gamma'_3\cong T'$ which is the double covering group of $\Gamma_3\cong A_4$. Apart from the three singlet representations $\mathbf{1}$, $\mathbf{1}'$, $\mathbf{1}''$ and a triplet representation $\mathbf{3}$ which are in common with these of $A_4$, the $T'$ group has three faithful independent doublets $\mathbf{2}$, $\mathbf{2}'$, $\mathbf{2}''$. The weight 1 modular forms of level 3 form a linear space $\mathcal{M}_{1}(\Gamma(3))$ of dimension 2, and they can be arranged into a doublet $\mathbf{2}$ of $T'$. We construct the multiplets of modular forms of weights 2, 3, 4, 5 and 6 from the tensor products of the weight 1 modular forms. It is remarkable that the odd weight modular forms transform in the two-dimensional irreducible representations $\mathbf{2}$, $\mathbf{2}'$, $\mathbf{2}''$ while the modular forms of even weight decompose as singlets and triplet under $T'$. Moreover, we see that the weights 2, 4, 6 modular form are exactly identical with those of~\cite{Feruglio:2017spp} up to irrelevant overall factors, the constraint satisfied by the weight 2 modular forms is quite obvious in our formalism. As a result, $T'$ is indistinguishable from $A_4$ if one works with even weight modular forms of level 3.

Finally we build a modular invariant model with $T'$ flavor symmetry. The neutrino masses are generated through the type I seesaw mechanism, and only two right-hand neutrinos are introduced and they are assigned to a $T'$ doublet. The structure of the model is rather simple, the superpotential is given in Eqs.~(\ref{eq:superpotential_We}, \ref{eq:superpotential_Wnu}), and the weight 3 modular forms are involved in the neutrino Yukawa couplings. The resulting neutrino and charged lepton mass matrices only depend on six free real parameters, the charged lepton masses and neutrino oscillation data can be accommodated very well, and the neutrino mass spectrum is determined to be inverted ordering.

It turns out that the doublet plus singlet assignment for the quark fields is more suitable to reproduce the hierarchial quark masses and mixing patterns~\cite{Feruglio:2007uu,Chen:2007afa,Ding:2008rj,Frampton:2008bz,Chen:2009gy,Girardi:2013sza}. It is appealing to extend the $T'$ modular symmetry to the quark sector, thus give a unified description of both quark and lepton mass hierarchies and flavor mixing. The odd weight modular forms provide new opportunity for building modular invariant models, it is worth further studying possible applications of the odd weight modular forms in understanding the flavor puzzle of SM. It is interesting to discuss other finite modular groups such as $\Gamma'_4$ and $\Gamma'_5$ and their phenomenological predictions for lepton masses and flavor mixing parameters.

\subsection*{Acknowledgements}

This  work  is  supported  by  the National Natural Science Foundation of China under Grant Nos 11522546 and 11835013.

\newpage

\section*{Appendix}

\setcounter{equation}{0}
\renewcommand{\theequation}{\thesection.\arabic{equation}}

\begin{appendix}
\section{\label{sec:Tp_group}Group Theory of $T^{\prime}$}

The homogeneous finite modular group $\Gamma'_3$ is isomorphic to $T'$ which is the double covering of the tetrahedral group $A_4$. It is well-known that $SU(2)$ is the double cover group of $SO(3)$, two different $SU(2)$ elements correspond to the same element of $SO(3)$. $A_4$ and $T'$ are subgroups of $SO(3)$ and $SU(2)$ respectively, and the $T'$ group can be regarded as the inverse image of the group $A_4$ under this map. The $T'$ group has 24 elements which can be generated by three generators $S$, $T$ and $\mathbb{R}$ fulfilling the following relations:
\begin{equation}
S^{2}=\mathbb{R},~~ (ST)^{3}=\mathbb{1},~~ T^{3}=\mathbb{1}, ~~\mathbb{R}^{2}=\mathbb{1},~~\mathbb{R}T = T\mathbb{R}\,.
\end{equation}
where $\mathbb{R}=\mathbb{1}$ in case of the odd-dimensional representation and
$\mathbb{R}=-\mathbb{1}$ for even-dimensional representations $\mathbf{2}$, $\mathbf{2}^{\prime}$ and $\mathbf{2}^{\prime \, \prime}$ such that $\mathbb{R}$ commutes with all elements of the group. The 24 elements of $T'$ group belong to 7 conjugacy classes:
\begin{eqnarray}
\nonumber 1 C_1:&&\,\mathbb{1}\,, \\
\nonumber 1 C_2:&&\,\mathbb{R}\,, \\
\nonumber 6 C_4:&&\,S,\,T^{-1}ST,\,TST^{-1},\,S\mathbb{R},\,T^{-1}ST\mathbb{R},\,TST^{-1}\mathbb{R} \,, \\
\nonumber 4 C_6:&&\,T\mathbb{R},\,TS\mathbb{R},\,ST\mathbb{R},\,T^{-1}ST^{-1}\mathbb{R} \,, \\
\nonumber 4 C_3:&&\,T^{-1},\,ST^{-1}\mathbb{R},\,T^{-1}S\mathbb{R},\,TST\mathbb{R} \,, \\
\nonumber 4 C'_3:&&\,T,\,TS,\,ST,\,T^{-1}ST^{-1} \,, \\
\label{T'conj_class} 4C'_6:&&\,ST^{-1},\,T^{-1}S,\,TST,\,T^{-1}\mathbb{R} \,,
\end{eqnarray}
\begin{table}[t!]
\centering
\begin{tabular}{|c||c|c|c||c|c|c||c|}
\hline\hline
&$\mathbf{1}$& $\mathbf{1'}$& $\mathbf{1''}$ & $\mathbf{2}$ & $\mathbf{2'}$ & $\mathbf{2''}$ &$\mathbf{3}$\\ \hline
$S$ & $1$ & $1$ & $1$ & $A_1$ & $A_1$ & $A_1$ &
$\frac{1}{3}\left(\begin{array}{ccc}
-1&2&2\\
2&-1&2\\
2&2&-1
\end{array}\right)$\rule[-4ex]{0pt}{10ex} \\ \hline
$T$ & $1$ & $\omega$ & $\omega^2$ & $A_2$ & $\omega A_2$ & $\omega^2 A_2$ &
$\left(
\begin{array}{ccc}
1~&0~&0\\
0~&\omega~&0\\
0~&0~&\omega^2
\end{array}\right)$\rule[-4ex]{0pt}{10ex} \\ \hline

$\mathbb{R}$ & 1 & 1 & 1 & $-\mathbb{1}_2$ & $-\mathbb{1}_2$ & $-\mathbb{1}_2$ &
$\mathbb{1}_3$ \rule[-1ex]{0pt}{1.5ex}\\ \hline\hline

\end{tabular}
\caption{\label{tab:irred}The representation matrices of the generators $S$, $T$ and $\mathbb{R}$ in different irreducible representations of $T'$ group, where $\omega=e^{i2\pi/3}=-\frac{1}{2}+i\frac{\sqrt{3}}{2}$ denotes a cubic root of unity, the matrices $A_1$ and $A_2$ are given in Eq.~\eqref{eq:A1A2}, and $\mathbb{1}_2$ and $\mathbb{1}_3$ are two-dimensional and three-dimensional unit matrices respectively. }
\end{table}
where $nC_k$ denotes a conjugacy class of $n$ elements which are of order $k$. Since the number of irreducible representation is equal to the number of conjugacy classes, the $T'$ group has seven inequivalent irreducible representations: three singlets $\mathbf{1}$, $\mathbf{1}'$ and $\mathbf{1}''$, three doublets $\mathbf{2}$, $\mathbf{2}'$ and $\mathbf{2}''$, and one triplet $\mathbf{3}$. The representations $\mathbf{1}'$, $\mathbf{1}''$ and $\mathbf{2}'$, $\mathbf{2}''$ are complex conjugated to each other. The odd dimensional representations $\mathbf{1}$, $\mathbf{1}'$, $\mathbf{1}''$ and $\mathbf{3}$ are representations of $A_4$. In these representations, two distinct $T'$ group elements correspond to the same matrix which represents the element in $A_4$. Consequently there is no way to distinguish the $T'$ group from $A_4$ when working with the odd dimensional representations. The $T'$ group as flavor symmetry for both quark and lepton has been discussed in the literature~\cite{Feruglio:2007uu,Chen:2007afa,Ding:2008rj,Frampton:2008bz,Chen:2009gy,Girardi:2013sza}.
In the present work we choose a basis similar to that of Refs.~\cite{Feruglio:2007uu,Ding:2008rj}, the explicit forms of the generators $S$, $T$ and $\mathbb{R}$ in each irreducible representations are listed in table~\ref{tab:irred} where we have used the matrices
\begin{equation}
\label{eq:A1A2}A_1=-\dfrac{1}{\sqrt{3}}\left(\begin{array}{cc}
 i & \sqrt2e^{i\pi/12} \\
 -\sqrt2e^{-i\pi/12} & -i \\
\end{array}\right)\,,\qquad
A_2=\left(\begin{array}{cc}
\omega & 0 \\
0 & 1 \\
\end{array}\right)\,.
\end{equation}
The Kronecker products between various irreducible representations of $T'$ are as follows:
\begin{align}
\begin{array}{l}
\mathbf{1}^a\otimes \mathbf{r}^b = \mathbf{r}^b \otimes \mathbf{1}^a= \mathbf{r}^{a+b~(\text{mod}~3)},~~~~~{\rm for}~~\mathbf{r} = \mathbf{1}, \mathbf{2}\,,\\
\mathbf{1}^a \otimes \mathbf{3} = \mathbf{3}\otimes \mathbf{1}^a=\mathbf{3}\,,\\
\mathbf{2}^a \otimes \mathbf{2}^b=\mathbf{3}\oplus \mathbf{1}^{a+b+1~(\text{mod}~3)}\,, \\
\mathbf{2}^a \otimes \mathbf{3} =\mathbf{3} \otimes \mathbf{2}^a = \mathbf{2}\oplus \mathbf{2'}\oplus \mathbf{2''}\,, \\
\mathbf{3} \otimes \mathbf{3} = \mathbf{3}_S \oplus \mathbf{3}_A \oplus \mathbf{1} \oplus \mathbf{1'} \oplus \mathbf{1''}\,,
\end{array}
\label{eq:mult}
\end{align}
where $a,b=0,\pm1$ and we have denoted $\mathbf{1}\equiv\mathbf{1}^0$, $\mathbf{1}'\equiv\mathbf{1}^{1}$, $\mathbf{1}''\equiv\mathbf{1}^{-1}$ for singlet representations and $\mathbf{2}\equiv\mathbf{2}^0$, $\mathbf{2}'\equiv\mathbf{2}^{1}$, $2''\equiv\mathbf{2}^{-1}$ for the doublet representations. We summarize the Clebsch-Gordan coefficients for the decomposition of product representations in our basis in table~\ref{tab:T'_CG}. We use $\alpha_i$ to denote the elements of the first representation, $\beta_i $ to indicate these of the second representation of the product.
\begin{table}[ht!]
\centering
\resizebox{1.1\textwidth}{!}{
\begin{tabular}{|c|c|c|c|c|c|}\hline\hline
\multicolumn{2}{|c|}{ ~$\mathbf{1}^a\otimes\mathbf{1}^b = \mathbf{1}^{a+b~(\text{mod}~3)}$ }
&\multicolumn{2}{|c|}{ $\mathbf{1}^a\otimes\mathbf{2}^b= \mathbf{2}^{a+b~(\text{mod}~3)}$ }
&\multicolumn{1}{|c|}{ ~~$\mathbf{1'}\otimes\mathbf{3} = \mathbf{3}$~~}
& \multicolumn{1}{|c|}{$\mathbf{1''}\otimes\mathbf{3} = \mathbf{3}$}  \\ \hline
 \multicolumn{2}{|c|}{} & \multicolumn{2}{|c|}{} & \multicolumn{1}{|c|}{} & \multicolumn{1}{|c|}{}\\[-0.1in]
 \multicolumn{2}{|c|}{$\mathbf{1}^{a+b~(\text{mod}~3)}\sim \alpha\beta $}
&
 \multicolumn{2}{|c|}{$\mathbf{2}^{a+b~(\text{mod}~3)}\sim
\left(\begin{array}{c} \alpha\beta_1\\
\alpha\beta_2 \\ \end{array}\right) $}
&
\multicolumn{1}{|c|}{~~$\mathbf{3} \sim
\left(\begin{array}{c} \alpha\beta_3\\
\alpha\beta_1 \\
\alpha\beta_2 \end{array}\right) $~~}
&
\multicolumn{1}{|c|}{$\mathbf{3} \sim
\left(\begin{array}{c} \alpha\beta_2\\
\alpha\beta_3 \\
\alpha\beta_1 \end{array}\right) $} \\ [+0.25in]\hline \hline
\multicolumn{2}{|c|}{$\mathbf{2}\otimes\mathbf{2}=\mathbf{2'}\otimes \mathbf{2''}=\mathbf{3}\oplus\mathbf{1'} $}
&\multicolumn{2}{|c|}{$\mathbf{2}\otimes\mathbf{2'}=\mathbf{2''}\otimes\mathbf{2''}=\mathbf{3}\oplus\mathbf{1''}$}
&\multicolumn{2}{|c|}{$\mathbf{2}\otimes\mathbf{2''}=\mathbf{2'}\otimes\mathbf{2'}=\mathbf{3}\oplus\mathbf{1} $~}\\ \hline
 \multicolumn{2}{|c|}{} & \multicolumn{2}{|c|}{} & \multicolumn{2}{|c|}{} \\[-0.1in]
\multicolumn{2}{|c|}{ $\mathbf{1'}\sim \alpha_1\beta_2-\alpha_2\beta_1$}
&
\multicolumn{2}{|c|}{$\mathbf{1''}\sim \alpha_1\beta_2-\alpha_2\beta_1$}
&
\multicolumn{2}{|c|}{$\mathbf{1} \sim \alpha_1\beta_2-\alpha_2\beta_1 $} \\
 \multicolumn{2}{|c|}{} & \multicolumn{2}{|c|}{} & \multicolumn{2}{|c|}{} \\[-0.1in]
\multicolumn{2}{|c|}{$\mathbf{3}\sim
\left(\begin{array}{c} e^{i \pi /6}\alpha_2\beta_2 \\
\frac{1}{\sqrt{2}}e^{i 7\pi /12}(\alpha_1\beta_2+\alpha_2\beta_1)  \\
\alpha_1\beta_1 \end{array}\right) $}
&\multicolumn{2}{|c|}{$\mathbf{3}\sim
 \left(\begin{array}{c} \alpha_1\beta_1\\
 e^{i \pi /6}\alpha_2\beta_2 \\
 \frac{1}{\sqrt{2}}e^{i 7\pi /12}(\alpha_1\beta_2+\alpha_2\beta_1) \end{array}\right) $}
&\multicolumn{2}{|c|}{$\mathbf{3}\sim
\left(\begin{array}{c}  \frac{1}{\sqrt{2}}e^{i 7\pi /12}(\alpha_1\beta_2+\alpha_2\beta_1)\\
 \alpha_1\beta_1\\
 e^{i \pi /6}\alpha_2\beta_2  \end{array}\right) $} \\[+0.25in] \hline \hline
\multicolumn{2}{|c|}{$\mathbf{2}\otimes\mathbf{3}=\mathbf{2}\oplus\mathbf{2'}\oplus\mathbf{2''} $}
&\multicolumn{2}{|c|}{$\mathbf{2'}\otimes\mathbf{3}=\mathbf{2}\oplus\mathbf{2'}\oplus\mathbf{2''}$}
&\multicolumn{2}{|c|}{$\mathbf{2''}\otimes\mathbf{3}=\mathbf{2}\oplus\mathbf{2'}\oplus\mathbf{2''}$} \\ \hline
\multicolumn{2}{|c|}{}& \multicolumn{2}{|c|}{}&\multicolumn{2}{|c|}{} \\[-0.1in]
\multicolumn{2}{|c|}{$\mathbf{2}\sim
 \left(\begin{array}{c} \alpha_1\beta_1-\sqrt{2}e^{i 7\pi /12}\alpha_2\beta_2 \\
 -\alpha_2\beta_1+\sqrt{2}e^{i 5\pi / 12}\alpha_1\beta_3 \end{array}\right) $}
&\multicolumn{2}{|c|}{$\mathbf{2}\sim
\left(\begin{array}{c} \alpha_1\beta_3-\sqrt{2}e^{i 7\pi /12}\alpha_2\beta_1 \\
-\alpha_2\beta_3+\sqrt{2}e^{i 5\pi / 12}\alpha_1\beta_2  \end{array}\right)$ }
&\multicolumn{2}{|c|}{$\mathbf{2}\sim
\left(\begin{array}{c} \alpha_1\beta_2-\sqrt{2}e^{i 7\pi /12}\alpha_2\beta_3 \\
-\alpha_2\beta_2+\sqrt{2}e^{i 5\pi / 12}\alpha_1\beta_1  \end{array}\right)$}  \\

\multicolumn{2}{|c|}{}& \multicolumn{2}{|c|}{}&\multicolumn{2}{|c|}{} \\[-0.1in]
\multicolumn{2}{|c|}{$\mathbf{2'}\sim
\left(\begin{array}{c} \alpha_1\beta_2-\sqrt{2}e^{i 7\pi /12}\alpha_2\beta_3 \\
-\alpha_2\beta_2+\sqrt{2}e^{i 5\pi / 12}\alpha_1\beta_1  \end{array}\right) $}
&\multicolumn{2}{|c|}{$\mathbf{2'}\sim
 \left(\begin{array}{c} \alpha_1\beta_1-\sqrt{2}e^{i 7\pi /12}\alpha_2\beta_2 \\
 -\alpha_2\beta_1+\sqrt{2}e^{i 5\pi / 12}\alpha_1\beta_3 \end{array}\right) $}
&\multicolumn{2}{|c|}{$\mathbf{2'}\sim
\left(\begin{array}{c} \alpha_1\beta_3-\sqrt{2}e^{i 7\pi /12}\alpha_2\beta_1 \\
-\alpha_2\beta_3+\sqrt{2}e^{i 5\pi / 12}\alpha_1\beta_2  \end{array}\right) $}\\
 \multicolumn{2}{|c|}{}& \multicolumn{2}{|c|}{}&\multicolumn{2}{|c|}{} \\[-0.1in]
\multicolumn{2}{|c|}{$\mathbf{2''}\sim
\left(\begin{array}{c} \alpha_1\beta_3-\sqrt{2}e^{i 7\pi /12}\alpha_2\beta_1 \\
-\alpha_2\beta_3+\sqrt{2}e^{i 5\pi / 12}\alpha_1\beta_2  \end{array}\right) $}
&\multicolumn{2}{|c|}{$\mathbf{2''}\sim
\left(\begin{array}{c} \alpha_1\beta_2-\sqrt{2}e^{i 7\pi /12}\alpha_2\beta_3 \\
-\alpha_2\beta_2+\sqrt{2}e^{i 5\pi / 12}\alpha_1\beta_1  \end{array}\right) $}
&\multicolumn{2}{|c|}{$\mathbf{2''}\sim
 \left(\begin{array}{c} \alpha_1\beta_1-\sqrt{2}e^{i 7\pi /12}\alpha_2\beta_2 \\
 -\alpha_2\beta_1+\sqrt{2}e^{i 5\pi / 12}\alpha_1\beta_3 \end{array}\right) $}  \\ [+0.15in] \hline \hline
\multicolumn{6}{|c|}{~$\mathbf{3}\otimes\mathbf{3}=\mathbf{3}_S\oplus\mathbf{3}_A\oplus\mathbf{1}\oplus\mathbf{1'}\oplus\mathbf{1''} ~$}\\ \hline
 \multicolumn{2}{|c|}{}& \multicolumn{2}{|c|}{}&\multicolumn{2}{|c|}{} \\[-0.1in]
\multicolumn{2}{|c|}{\multirow{3}{*}{$\mathbf{3}_S\sim
 \left(\begin{array}{c} 2\alpha_1\beta_1 - \alpha_2\beta_3 - \alpha_3\beta_2 \\
 2\alpha_3\beta_3 - \alpha_1\beta_2 - \alpha_2\beta_1  \\
 2\alpha_2\beta_2 - \alpha_1\beta_3 - \alpha_3\beta_1 \end{array}\right) $} }
&\multicolumn{2}{|c|}{\multirow{3}{*}{$\mathbf{3}_A\sim
 \left(\begin{array}{c} \alpha_2\beta_3 - \alpha_3\beta_2 \\
\alpha_1\beta_2 - \alpha_2\beta_1  \\
 \alpha_3\beta_1 - \alpha_1\beta_3 \end{array}\right)$} }
&  \multicolumn{2}{|c|}{$\mathbf{1} \sim \alpha_1\beta_1 + \alpha_2\beta_3 + \alpha_3\beta_2$ }\\
\multicolumn{2}{|c|}{}&\multicolumn{2}{|c|}{} & \multicolumn{2}{|c|}{$\mathbf{1'} \sim \alpha_3\beta_3 + \alpha_1\beta_2 + \alpha_2\beta_1$ }\\
\multicolumn{2}{|c|}{}&\multicolumn{2}{|c|}{}& \multicolumn{2}{|c|}{$\mathbf{1''} \sim \alpha_2\beta_2 + \alpha_1\beta_3 + \alpha_3\beta_1$ }\\[+0.05in] \hline \hline
\end{tabular} }
\caption{\label{tab:T'_CG}The Kronecker products and Clebsch-Gordan coefficients of the $T'$ group. The tensor product of two irreducible representations $\mathbf{R_1}$ and $\mathbf{R_2}$ is reported in the form of $\mathbf{R_1}\otimes \mathbf{R_2}$. The elements of $\mathbf{R}_1$ and $\mathbf{R}_2$ are labeled as $\alpha_i$ and $\beta_i$ respectively.}
\end{table}

\end{appendix}

\newpage

\providecommand{\href}[2]{#2}\begingroup\raggedright\endgroup

\end{document}